\documentclass[aps,showpacs,a4paper,floatfix,twocolumn,prc,amsmath,amssymb]{revtex4-1}
\usepackage{amsmath}
\usepackage{graphicx}
\usepackage{ulem}
\usepackage{morefloats}
\usepackage{rotating}
\usepackage{relsize}
\usepackage{color}
\usepackage{floatflt,epsfig}
\usepackage{ulem}
\usepackage{natbib}
\usepackage{hyperref}
\usepackage{filecontents}

\begin{document}

\title{Pasta phases within the QMC model}

\author{Guilherme Grams $^1$, Alexandre M. Santos $^1$, Prafulla
K. Panda $^2$,  Constan\c ca Provid\^encia $^3$, D\'ebora P. Menezes $^1$}

\affiliation{$^1$Depto de F\'isica, CFM, Universidade Federal de Santa Catarina, Brazil\\
$^2$Department of Physics, Utkal University, Bhubaneswar-751 004, India\\
$^3$CFisUC, Department of Physics, University of Coimbra, P-3004-516 Coimbra, Portugal}
\begin{abstract}
In this work the low density regions of nuclear and neutron star
matter are studied. The search for the existence of pasta phases 
in this region is performed within the context of the quark-meson
coupling (QMC) model, which incorporates quark
degrees of freedom. 
Fixed proton fractions are considered, as well as nuclear matter in
beta equilibrium at zero temperature. We discuss the recent 
attempts to better understand the surface energy in the coexistence
phases regime and we present results that show the existence of the
pasta phases subject to some choices of the surface energy coefficient.
We also analyze the influence of the  nuclear pasta on some neutron
star properties.
The equation of state containing the pasta phase will be part of a
complete grid for future use in supernova simulations.
\end{abstract}


\maketitle

\section{Introduction}

At very low nuclear matter density, a competition between
the strong and the electromagnetic interactions takes place \cite{Avancini-08, xubao}, leading to
a configuration in which its free energy per particle may be lower than
the corresponding to the homogeneous phase at the same density. The so-called 
{\it pasta phases} are therefore the preferred shapes of some systems
at these densities \cite{Ravenhall83,horo04,horo05,maruyama}.
These structures look like droplets, bubbles, rods, tubes and slabs \cite{Ravenhall83}, 
and are expected to exist \cite{pp98, NScrust_book} both in the 
crust of neutron stars (zero temperature, very low proton fraction, matter 
in $\beta$- equilibrium) and in supernova (finite temperature, proton fraction
around 0.3).

From the analysis of glitches, the authors of \cite{Link-99} have related the fraction of the moment 
of inertia contained in the crust of the Vela pulsar with the mass and the radius of 
the neutron star and the pressure and density at the crust-core interface. From realistic 
EoS they have obtained an expected range of values for the pressure at the 
inner edge of the crust and therefore also a relation between the radius and mass
of the pulsar. This work shows the importance of 
understanding the exact density limits of the pasta phases and its consequences 
on the choice of appropriate equations of state.
More recently, the existence of the pasta phase in the neutron star crust was shown to
considerably alter the neutrino mean-free paths and its diffusion
coefficients as compared with the homogeneous matter results. The
consequent differences in neutrino opacities certainly influence the
Kelvin-Helmholtz phase of the star evolution \cite{alloy-2011,Furtado16}.

On the other hand, due to well known observational difficulties,
simulations of core-collapse supernova have played an important role
in the study of supernovae explosions and the evolution of their possible remnants.
Hence, obtaining appropriate equations of state (EoS) for core-collapse 
supernova simulations has been a very challenging task. 
For this class of EoS one needs a grid of thermodynamic quantities
with densities ranging  from $10^{5}$ to more than 
$10^{15}$ g.cm$^{-3}$, proton fractions up to about $0.6$ and
temperatures varying from zero to more than $100$ MeV. 
So far, in the models used for the
obtention of a complete grid with the aim of being tested in supernova simulations,
inhomogeneous matter believed to be present at low densities, has only
been considered with the inclusion of clusters
\cite{shen98,shen11,hempel,sfhxandsfho,banikhempel}.
 However, according to the works of \cite{Sonoda07,Sonoda08} 
the pasta phases may form $10-20 \% $ of the mass of the supernova
core, therefore its role should not be disregarded. 
The pasta phases have been studied in the context of several models 
\cite{Avancini-08,Sonoda08,Avancini-09,Watanabe09,helena12} 
 and all of them predict its existence under 
the conditions expected to be found in the inner crust of compact objects , 
although the profiles show that they vary in many aspects
\cite{Avancini-09, NScrust_book}.

The quark meson coupling (QMC) model \cite{guichon, saito, saito95}
describes nuclear matter as a system of non-overlapping MIT-like bags,
interacting with each other by interchanging meson fields. Hence, it
contains more fundamental degrees of freedom than the usual quantum
hadrodynamic models, so far used to the study of the pasta phases \cite{Avancini-08,
Avancini-09, alloy-2011, helena16}. With the aim of constructing a complete grid for supernova simulations, 
a preliminary work at zero temperature, $\rho=10^{14}-10^{16}$
g.cm$^{-3}$ and $Y_p=0-0.65$ was done \cite{grams16} and revealed that the QMC is a 
promising model.

In the present work, we study the possible existence of the pasta
structures within the QMC model at zero temperature and its dependence on
the surface energy coefficient. The work is organized as follows: In
Sec. \ref{qmc} the QMC model is briefly reviewed and the method of the
coexisting phases used to build the pasta phase is presented in
Sec. \ref{cophases}, where a detailed study of the surface tension
coefficient is performed. In Sec. \ref{results} we present our results and
draw the conclusions. In the last Section, we make some final remarks.

%-------------------------------------------------------------------------------
\section{The quark-meson coupling model}
\label{qmc}

In the QMC model, the nucleon in nuclear medium is assumed to be a
static spherical MIT bag in which quarks interact with the scalar ($\sigma$)
and vector ($\omega$, $\rho$) fields, and those
are treated as classical fields in the mean field
approximation (MFA) \cite{guichon}.
The quark field, $\psi_{q_{N}}$, inside the bag then
satisfies the equation of motion:
\begin{eqnarray}
\left[i\,\rlap{/}\partial \right.&-&(m_q^0-g_\sigma^q\,)-g_\omega^q\, \omega\,\gamma^0\nonumber \\
&+&\left.  \frac{1}{2} g^q_\rho \tau_z \rho_{03}\gamma^0\right]
\,\psi_{q_{N}}(x)=0\ , \quad  q=u,d
\label{eq-motion}
\end{eqnarray}
where $m_q^0$ is the current quark mass, and $g_\sigma^q$, $g_\omega^q$ and $g_\rho^q$
denote the quark-meson coupling constants. The normalized ground state for a quark in the bag 
is given by
\begin{eqnarray}
\psi_{q_{N}}({\bf r}, t) &=& {\cal N}_{q_{N}} \exp 
\left(-i\epsilon_{q_{N}} t/R_N \right) \nonumber \\
&\times& \left(
\begin{array}{c}
  j_{0_{N}}\left(x_{q_{N}} r/R_N\right)\\
i\beta_{q_{N}} \vec{\sigma} \cdot \hat r j_{1_{N}}\left(x_{q_{N}} r/R_N\right)
\end{array}\right)
 \frac{\chi_q}{\sqrt{4\pi}} ~,
\end{eqnarray}
where
\begin{equation}
\epsilon_{q_{N}}=\Omega_{q_{N}}+R_N\left(g_\omega^q\, \omega+
 \frac{1}{2} g^q_\rho \tau_z \rho_{03} \right) ,
\end{equation}
and,
\begin{equation}
\beta_{q_{N}}=\sqrt{\frac{\Omega_{q_{N}}-R_N\, m_q^*}{\Omega_{q_{N}}\, +R_N\, m_q^* }}\ ,
\end{equation}
with the normalization factor given by
\begin{equation}
{\cal N}_{q_{N}}^{-2} = 2R_N^3 j_0^2(x_q)\left[\Omega_q(\Omega_q-1)
+ R_N m_q^*/2 \right] \Big/ x_q^2 ~,
\end{equation}
where $\Omega_{q_{N}}\equiv \sqrt{x_{q_{N}}^2+(R_N\, m_q^*)^2}$,
$m_q^*=m_q^0-g_\sigma^q\, \sigma$, $R_N$ is the
bag radius of nucleon $N$ and $\chi_q$ is the quark spinor. The bag eigenvalue for nucleon $N$, $x_{q_{N}}$, is determined by the
boundary condition at the bag surface
\begin{equation}
j_{0_{N}}(x_{q_{N}})=\beta_{q_{N}}\, j_{1_{N}}(x_{q_{N}})\ .
\label{bun-con}
\end{equation}

The energy of a static bag describing nucleon $N$ consisting of three quarks in ground state
is expressed as
\begin{equation}
E^{\rm bag}_N=\sum_q n_q \, \frac{\Omega_{q_{N}}}{R_N}-\frac{Z_N}{R_N}
+\frac{4}{3}\,  \pi \, R_N^3\,  B_N\ ,
\label{ebag}
\end{equation}
where $Z_N$ is a parameter which accounts for zero-point motion
of nucleon $N$ and $B_N$ is the bag constant.
The set of parameters used in the present work is determined by enforcing 
stability of the nucleon (here, the ``bag''), much like in \cite{alex09}, 
so there is a single value for 
proton and neutron masses. The effective mass of a nucleon bag at rest
is taken to be $M_N^*=E_N^{\rm bag}.$

The equilibrium condition for the bag is obtained by
minimizing the effective mass, $M_N^*$ with respect to the bag radius
\begin{equation}
\frac{d\, M_N^*}{d\, R_N^*} = 0,\,\,\;\;\; N=p,n ,
\label{balance}
\end{equation}
By fixing the bag radius $R_N=0.6$ fm and the bare nucleon mass $M=939$ MeV the unknowns $Z_N=4.0050668$ and 
$B^{1/4}_N=210.85$MeV are then obtained.
Furthermore, the desired values of $ B/A \equiv   \epsilon/\rho - M = -15.7$~MeV at saturation
$n=n_0=0.15$~fm$^{-3}$, are achieved by setting $g_\sigma^q=5.9810$, $g_{\omega}=8.9817$, 
$g_\rho = 8.6510$,
where $g_\omega =3g^q_\omega$ and $g_\rho =g^q_\rho$. The meson masses are $m_{\sigma}=550$ MeV,
$m_{\omega}=783$ MeV and $m_{\rho}=770$ MeV. With this
parameterization, some of the bulk properties at saturation density are
the compressibility, the symmetry energy and the slope of the symmetry
energy, whose values can be seen in table \ref{tab:bulk}.
These numbers are very close to the most accepted
values (see \cite{dutra14}, for instance) and $J$ and $L_0$ can be
easily controlled by the inclusion of a $\omega-\rho$ interaction, as discussed
in \cite{EPJA_2014,Prafulla_2012,Cavagnoli_2011}. The larger the value of this
interaction, the lower the values of the symmetry energy and its slope. 
Other parameterizations are also possible. Of particular interest is
the modified QMC model, where the parameters are adjusted so that the
constituent quarks are confined to a flavour-independent potential
where pionic and gluonic corrections are taken into account \cite{rnm,hss}.
These studies will be performed in future investigations. Within the parameterization we have chosen, the
total energy density of the nuclear matter reads 
\begin{eqnarray}
\varepsilon = \frac{1}{2}m^{2}_{\sigma}\sigma+\frac{1}{2}m^{2}_{\omega}\omega^{2}_{0}+\frac{1}{2}m^{2}_{\rho}\rho^{2}_{03}\nonumber \\
+\sum_{N} \frac{1}{\pi^{2}}\int^{k_{N}}_{0}k^{2}dk[k^{2}+M^{*2}_{N}]^{1/2},
\label{energ dens}
\end{eqnarray}
and the pressure is,
%\bigskip
%
\begin{eqnarray}
p = -\frac{1}{2}m^{2}_{\sigma}\sigma+\frac{1}{2}m^{2}_{\omega}\omega^{2}_{0}+\frac{1}{2}m^{2}_{\rho}\rho^{2}_{03}\nonumber \\
+\sum_{N} \frac{1}{\pi^{2}}\int^{k_{N}}_{0}k^{4}dk/[k^{2}+M^{*2}_{N}]^{1/2}.
\label{press}
\end{eqnarray}

The vector mean field $ \omega_0 $ and $ \rho_{03} $ are determined through
\begin{equation}
\omega_0 =\frac{g_\omega (n_p +n_n)}{m^{2}_{\omega}}, \;
\rho_{03}=\frac{g_\rho (n_p -n_n)}{2 m^{2}_{\rho}},
\end{equation}
where
\begin{equation}
n_B=n_p+n_n = \sum_N \frac{2 k_{N}^3}{3 \pi ^2}, \quad N=p,n.
\end{equation}
is the baryon density. 

Finally, the mean field $\sigma$ is fixed by imposing that
\begin{equation}
\frac{\partial \varepsilon}{\partial \sigma}=0.
\end{equation}
%

%%---------------------tabela bulk qmc-----------------------------------
\begin{table}
\begin{tabular}{lclcccc}
\hline
Model   &$B/A$ & $n_{0}$   & $M^*/M$  & $J$&  $L_0$   &   K     \\
        & (MeV)&(fm$^{-3}$)&          & (MeV)             &  (MeV) & (MeV)   \\
\hline
QMC    	&-15.7 &0.150      & 0.77     & 34.5              &  90   & 295     \\
\hline
\end{tabular}
\caption{Nuclear matter bulk properties obtained with the QMC model. All quantities are taken at saturation.} 
\label{tab:bulk}
\end{table}

As mentioned in the Introduction, our interest lies on matter at fixed
proton fraction given by $Y_p=n_p/n_B$ as well as in stellar matter 
in  $\beta$-equilibrium conditions, which for the system made up of protons,
neutrons and electrons are: 
\begin{equation}
\mu_p=\mu_n-\mu_e.
\end{equation}
Charge neutrality requires that
\begin{equation}
n_p=n_e.
\end{equation}

In this article we work with the low density regions of the neutron
stars and in this region muons are not present. 

%--------------------------------------------------------------------------------
\section{Coexisting phases approximation}
\label{cophases}

In this approximation matter is organized in regions of lower
density, generally with a neutron gas in the background and
regions of higher density. For a given
total density $n_B$ and proton fraction $Y_p$, the pasta 
structures are built with different geometrical forms.
The forms are usually called:  sphere (bubble),  cylinder (tube), and slab, in three, two,
and one dimensions, respectively.
This is achieved by calculating the density and the proton fraction 
of the pasta and of the background gas from the Gibbs conditions,
that impose that both phases have the same pressure and proton
and neutron chemical potentials, so that the following 
equations must be solved simultaneously:

\begin{equation}
P^I=P^{II},
\label{p1e2}
\end{equation}
\begin{equation}
\mu_p^I=\mu_p^{II},
\end{equation}
\begin{equation}
\mu_n^I=\mu_n^{II},
\end{equation}
\begin{equation}
n_{p}=n_BY_{p}=f\;n_p^{I}+(1-f)n_p^{II},
\end{equation}
where $I$ ($II$) label the high-(low-)density phase, $n_p$
is the global proton density, $f$ is the volume fraction
of phase $I$,

\begin{equation}
f=\frac{n_B-n_B^{II}}{n_B^I-n_B^{II}}.
\end{equation} 

If stellar matter is considered, the above equations are slightly
altered in such a way that:

\begin{equation}
\mu^{I}_n=\mu^{II}_n,
\end{equation}
\begin{equation}
\mu^{I}_e=\mu^{II}_e
\end{equation}
and
\begin{equation}
f(n^{I}_p-n^{I}_e)+(1-f)(n^{II}_p-n^{II}_e)=0.
\end{equation}
 along with eq. (\ref{p1e2}).
Here the density of electrons is no longer uniform as in the fixed proton fraction case. It appears
as the solution of the above equation.

After the lowest energy state is achieved, the surface and Coulomb terms are added to the total energy density of the system, which is given by

\begin{eqnarray}
\varepsilon=f\varepsilon^I+(1-f)\varepsilon^{II}+\varepsilon_e+\varepsilon_{surf}+\varepsilon_{Coul}.
\end{eqnarray}
 By minimizing the sum $\varepsilon_{surf}+\varepsilon_{Coul}$
with respect to the size of the droplet/bubble, rod/tube or slab we get \cite{maruyama} 
$\varepsilon_{surf}=2 \varepsilon_{Coul}$ where
\begin{equation}
\varepsilon_{Coul}=\frac{2\alpha}{4^{2/3}}(e^{2}\pi \Phi)^{1/3}\left[ {\cal S} D(n^{I}_p -n^{II}_p)\right] ^{2/3},
\end{equation}
where $\alpha=f$ for droplets, rods and slabs, and $\alpha=1-f$ for
tubes and bubbles. 
${\cal S}$ is the surface tension discussed in the next subsection and
$\Phi$ is given by
\begin{eqnarray}
\Phi=\left\lbrace\begin{array}{c}
\left(\frac{2-D \alpha^{1-2/D}}{D-2}+\alpha\right)\frac{1}{D+2}, \quad
                  D=1,3 \\
\frac{\alpha-1-\ln \alpha}{D+2},  \quad D=2 \quad .
\end{array} \right. 
\end{eqnarray}

As we are treating only the low density region, the nucleon effective mass $M_N^*$ can be parametrized as
\begin{equation}
M_N^*=M_N-g_{\sigma N}(\sigma)\sigma
\end{equation}
with
\begin{equation}
g_{\sigma N}(\sigma)=(1+\frac{b}{2} \sigma +\frac{c}{3} \sigma^2)g_{\sigma N}
\end{equation}
where $g_{\sigma N}=3g_\sigma^qS_N(0)=8.6157$, $b=0.000722089$
MeV$^{-1}$ and $c=1.17509\times 10^{-7}$ MeV$^{-2}$. Notice that these
values are valid only for this specific parameterization.

Before we proceed to the discussion of the surface tension coefficient,
it is important to point out that the coexistence phase (CP) method does not take into
account the Coulomb interaction and finite-size
effects in a selfconsistent way. An alternative prescription within the compressible liquid
drop (CLD) model incorporates these important effects by minimizing the
total free energy, where surface and Coulomb terms are explicitly
included \cite{cld_shen} self-consistently.
The resulting pressure and proton chemical potential equilibrium 
conditions are slightly different from the ones above. The differences
between both prescriptions (CP and CLD) can be easily seen in
\cite{helena15} and the resulting pasta properties differ at very low
densities \cite{cld_shen, helena15}, generally lower than $10^{-3}$
 fm$^{-3}$ when the matching to the outer crust EoS is performed. 
As will be shown next, our calculation depends also on a
free parameter, that is fitted according to accepted values of
the surface tension.

\subsection{The surface tension coefficient}

In order to achieve a numerical value for the surface tension coefficient,
the geometrical approach introduced in \cite{Randrup09} is used next. In \cite{Pinto12},
this method was used to compute the surface tension in quark matter
but recently it  was also used to obtain
the surface tension coefficient for hadronic matter \cite{helena16}.
The main ideas are also discussed next.\\

\begin{figure}[!]
\includegraphics[width=9.cm,angle=0]{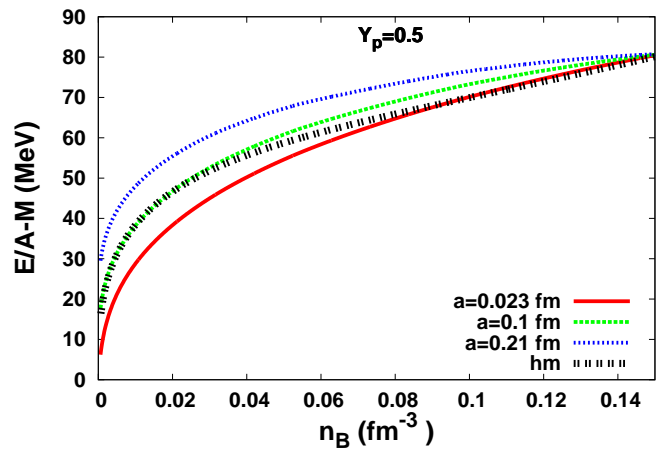}
\caption{Energy per baryon as a function of the baryon density proton
  fraction 0.5 and different choices of $a$.
 hm stands for homogeneous $npe$ matter.}
\label{comp}
\end{figure}

The surface tension coefficient, ${\cal S}$, which measures the
energy per unit area necessary to create a planar interface between the
two phases is given by
\begin{equation}
{\cal S}=\frac{a}{n_g}\sqrt{2\varepsilon_g}\int_{n_1}^{n_2}
\sqrt{\Delta\varepsilon} dn,
\end{equation}
where $n_g=\frac{n_1+n_2}{2}$, $\varepsilon_g=\frac{\varepsilon(n_1)+\varepsilon(n_2)}{2}$
and $\Delta \varepsilon=\varepsilon_{hm}-\varepsilon_{nhm}$ is the
difference between the energy density of the homogeneous and the
non-homogeneous matter. These energy densities are fitted to a functional
form given by $\varepsilon=a n^2+b n+c$, where $n_1$ and $n_2$
are the two coexistence baryonic density points. In this geometrical
approach, the width of the interface region and the
magnitude of ${\cal S}$ are controlled by the adjustable parameter $a$.
In \cite{Pinto12} the authors used $a=1/m_{\sigma}=0.33$ fm where $m_\sigma=600$ MeV is the mass of the $\sigma$ meson, a natural 
scale for quark matter. As we are treating hadronic matter in the
present work, our initial guess was $a=1/M_N=0.21$ fm, where $M_N=939$
MeV is the nucleon mass. Another attempt followed the recipe used to find the surface tension of
hadronic matter in \cite{helena16} with an extended version of the
Nambu-Jona-Lasino model, where $a=0.1$ fm was adopted to reproduce the
value of the surface tension coefficient for the NL3 model \cite{nl3}
within a Thomas-Fermi calculation \cite{avancini10}. 
Our final choice was  $a=0.023$ fm,  so that the value
${\cal S}=1.123$ MeV.fm$^{-2}$ for $Y_p=0.5$ was reproduced as in 
\cite{avancini10,avancini12,helena16}.
In Figure \ref{comp} we compare the three choices of $a$ in the search for 
the pasta phases, i.e., $a=0.023$ fm, $a=0.1$ fm and $a=1/M_N=0.21$ fm, 
for $Y_p=0.5$. 
We can see that there is a larger region of the pasta phase for $a=0.023$ fm. In fact,
for $Y_P=0.5$ no pasta phases were found with $a=0.1$ fm, neither with
$a=0.21$ fm. Therefore, we have chosen 
$a=0.023$ fm to be used throughout our calculations. 

It is important to stress that the surface tension coefficient varies with the isospin for  a given value of $a$.
In Table \ref{tab:sigs} we show the values of ${\cal S}$ for five different proton fractions.
In order to obtain the pasta phases in $\beta$-equilibrium matter, we fitted this values of 
${\cal S}$ to a functional of the form ${\cal S}=d+ex+fx^2$, with 
$d=-0.0389543$, $e=2.45143$ and $f=-0.209076$, where $x$ is the global
proton fraction. We note that there are some  works where the proton
faction used in the calculation of the surface tension is the one of the denser phase.

%%%%%%%%%%%%%%%%%%%figuras parametro a%%%%%%%%%%%%%%%%%%%%%%%%%%%%%%%5

\begin{figure}[!]
\begin{tabular}{ll}
\includegraphics[width=9.cm,angle=0]{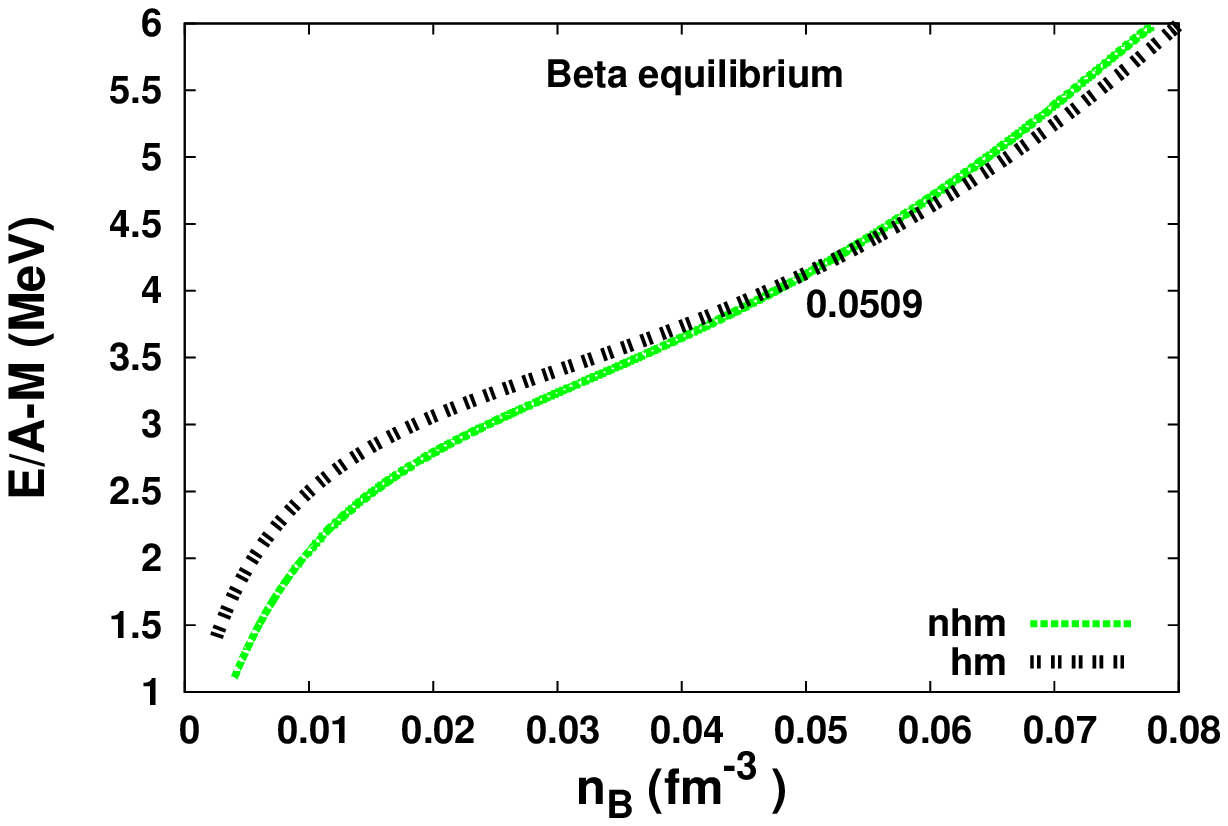} \\
\includegraphics[width=9.cm,angle=0]{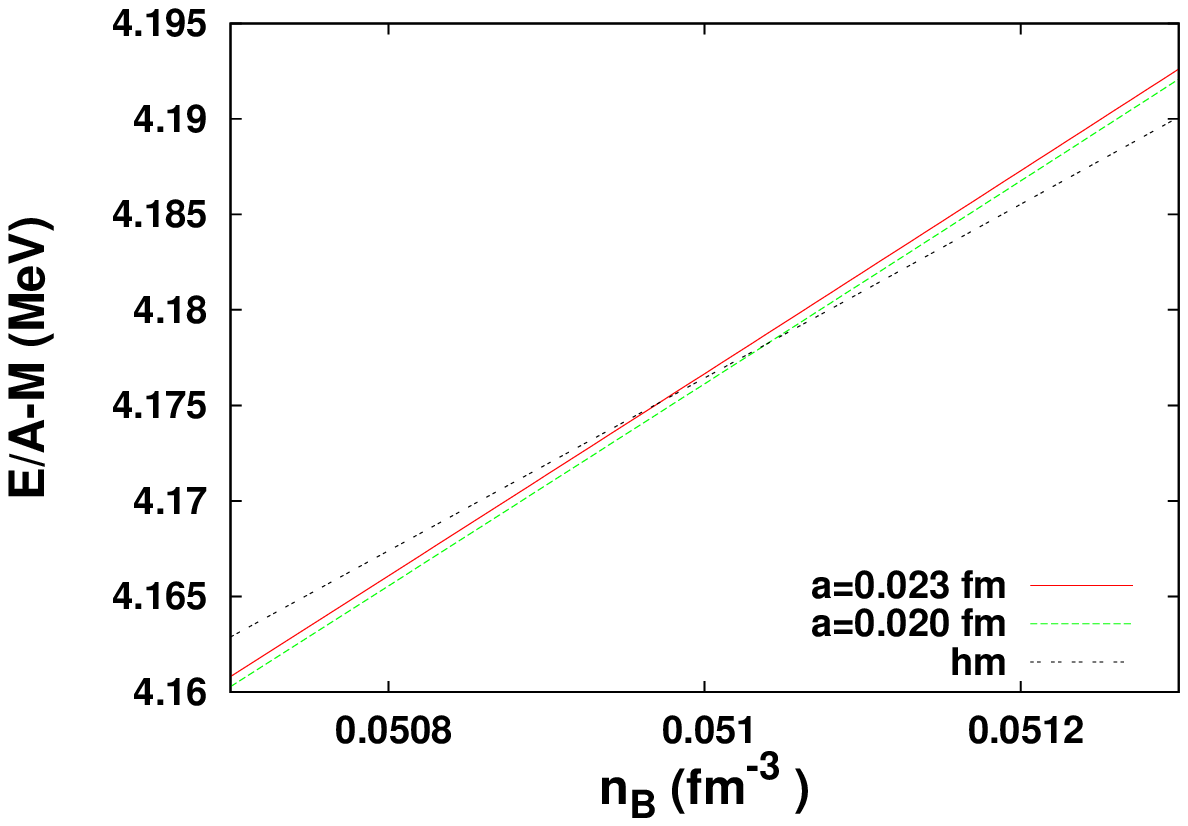}
\end{tabular}
\caption{Free energy per particle obtained for matter in $\beta$-equilibrium. The bottom figure shows a zoom of the transition
region from the pasta phases to homogeneous matter with two different choices 
of the parameter $a$.}
\label{figcomp}
\end{figure}
%%%%%%%%%%%%%%%%%%%%%%%%%%%%%%%%%%%%%%%%%%%%%%%%%%%%%%%%%%%%%%%%%%%%%%%%

\begin{table}
\centering
\begin{tabular}{llllllcc}

\hline
$Y_p$   \; &\;\; ${\cal S}$  (MeV.fm$^{-2}$)  \\
\hline
0.1     \; & \; 0.21 \\
0.2     \; & \; 0.42 \\
0.3     \; & \; 0.69  \\
0.4     \; & \; 0.93   \\ 
\hline
0.5     \; & \;  1.12     \\
\hline
\end{tabular}
\caption{Surface tension coefficient for different proton fractions.} 
\label{tab:sigs}
\end{table}

\begin{figure}[!]
\includegraphics[width=9.cm,angle=0]{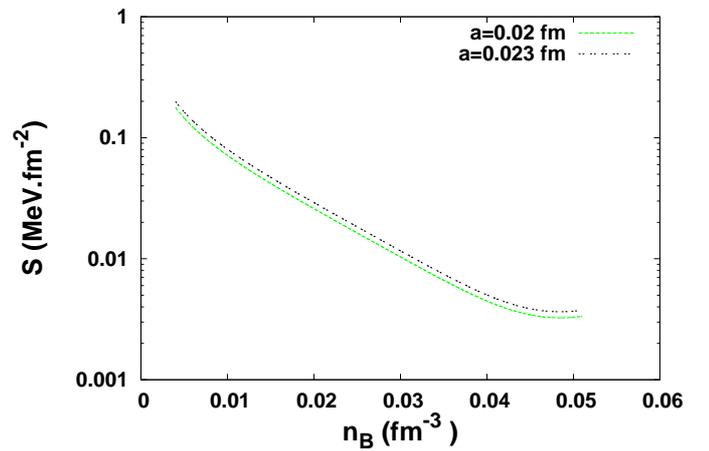}
\caption{Surface tension coefficient for $\beta$-equilibrium as function
of the baryon density for two choices of $a$.}
\label{sfbeta}
\end{figure}

Previous works \cite{maruyama,Avancini-08,Avancini-09,avancini10,avancini12,helena16}
have shown that the surface tension at zero temperature, not only
varies with the proton fraction, but present 
values in between 1.0 and 1.2 MeV.fm$^{-2}$ for $Y_p=0.5$ (see Fig. 5
in \cite{EPJA_2014}, for instance). If we constrain the parameter $a$
so that these values are reproduced, we obtain $a=0.020 - 0.025$ fm. 
The value $a=0.020$ fm yields ${\cal S}=1.0$ MeV.fm$^{-2}$
and $a=0.025$ fm results in ${\cal S}=1.2$ MeV.fm$^{-2}$ for $Y_p=0.5$. 
Choosing $a=0.025$ would increase the surface energy in 8.7\% having 
a very small effect on the crust-core transition of beta-equilibrium matter.
Therefore we proceed with the 
comparison of the results obtained with $a=0.020$ and with $a=0.023$,
which entails ${\cal S}=1.123$ MeV.fm$^{-2}$ for 
$Y_p=0.5$.
  We then compare our results with those two values of $a$ in
Fig. \ref{figcomp}, where the free energy per baryon in function of the
baryon density for matter in $\beta$-equilibrium is shown. We can see
that our results are practically independent of
$a$, as far as a reasonable value for the surface tension is used.
In Table \ref{tab:sigs2} we compare the surface tension coefficient
${\cal S}$ and the transition density $\rho_t$ for the two values of $a$.
We see that $\rho_t$ is practically independent of $a$ in the range
[0.020, 0.023], not only
for $\beta$-equilibrium matter but also for matter with fixed proton
fractions.
The surface tension coefficient as a function of the baryon density
is displayed in Fig. \ref{sfbeta} for matter in $\beta$-equilibrium, 
from where we note that ${\cal S}$ decreases  with the density.
We can see that the surface tension coefficient is only slightly larger for
$a=0.023$ fm  both from Table \ref{tab:sigs2} and Fig. \ref{sfbeta}.

\begin{table}
\centering
\begin{tabular}{lllllll}

\hline
$Y_p$    &  ${\cal S}$     &  $a$   & $\rho_t$ \\
         & (MeV.fm$^{-2}$) & (fm) &(fm$^{-3}$)\\
\hline
0.5 &  1.0       & 0.020  & 0.100  \\
0.5 &  1.12      & 0.023 & 0.097 \\
0.3 & 0.60      & 0.020  & 0.094 \\
0.3 & 0.69      & 0.023 & 0.093 \\
0.1 & 0.20       & 0.020  & 0.064 \\
0.1 & 0.21      & 0.023 & 0.063 \\
$\beta$-eq &  plot  & 0.020  & 0.051 \\ 
$\beta$-eq &  plot  & 0.023 & 0.051 \\ 
\hline
\end{tabular}
\caption{Surface tension coefficient for different proton
  fractions and related $a$ values. $\rho_t$ is the transition density
that separates the pasta from the homogeneous phase.} 
\label{tab:sigs2}
\end{table}

%%%%%%%%%%%%%%%%%%%figura sigma equilibrio beta%%%%%%%%%%%%%%%%%5
%%%%%%%%%%%%%%%%%%%%%%%%%%%%%%%%%%%%%%%%%%%%%%%%%%%%%%%%%%%%%%%5

%\newpage
\section{Results and Conclusions}
\label{results}

Finally, we present our results for the pasta phases obtained with
the QMC model at zero temperature, within the coexisting phases approximation.
We would like to remark that pasta is only predicted when its
free energy per baryon is lower than the homogeneous $npe$
(neutron-proton-electron) matter.

\begin{figure}[!]
\begin{tabular}{lll}
\includegraphics[width=8.cm,angle=0]{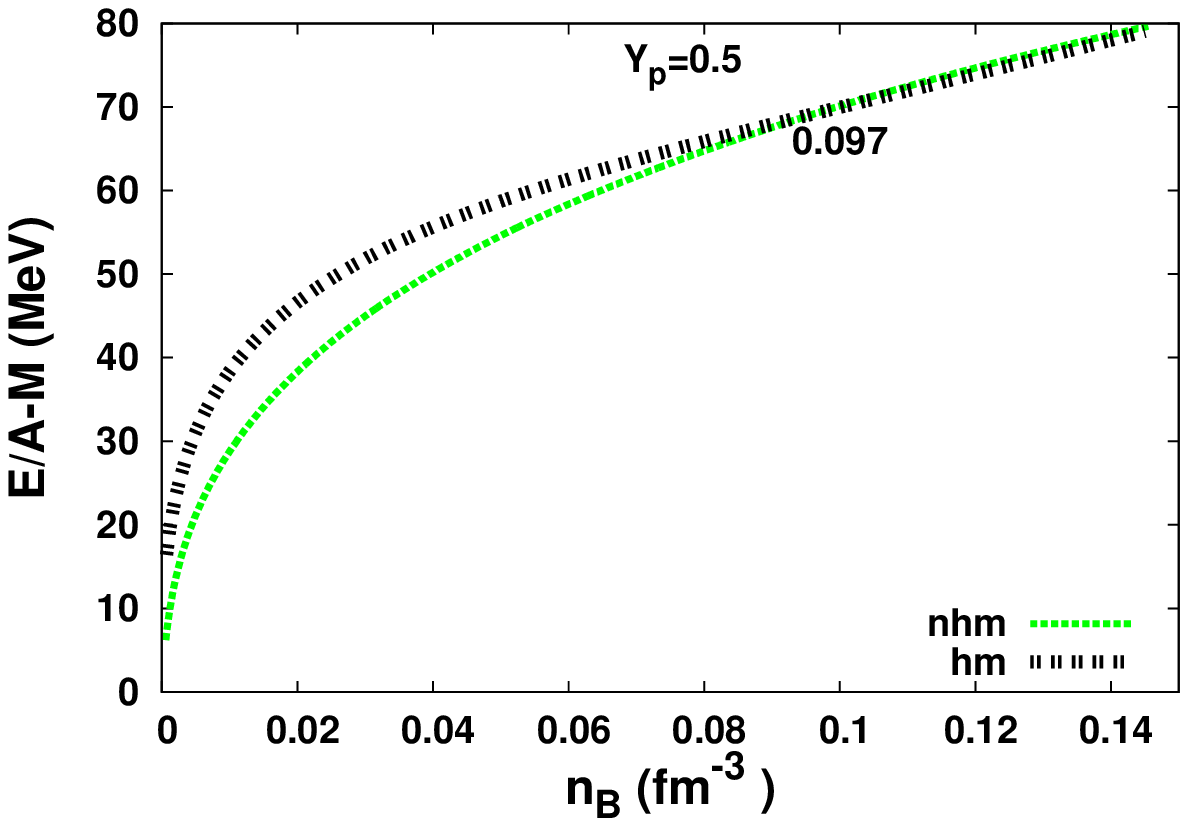}\\
\includegraphics[width=8.cm,angle=0]{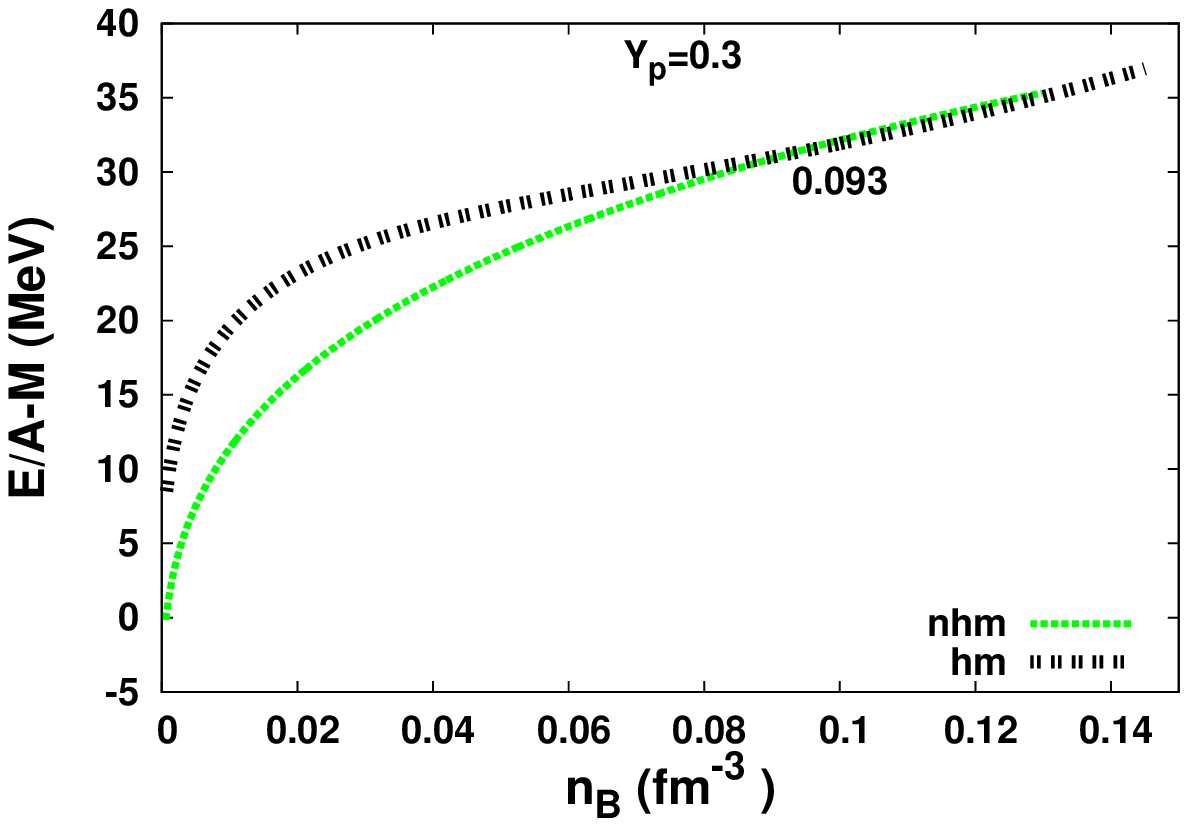}
\end{tabular}
\caption{Free energy per baryon as function of the baryon density with
  (a) the proton fraction 0.5, (b) proton fraction 0.3 
%and (c) beta equilibrium. 
hm=homogeneous matter and nhm=non-homogeneous matter.}
\label{3fig}
\end{figure}

\begin{figure}[!]
\begin{tabular}{l}
\includegraphics[width=8.cm,angle=0]{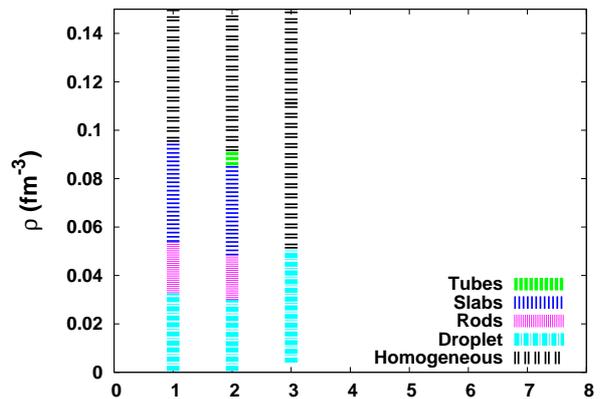}
\end{tabular}
\caption{Phase diagrams at T=0 obtained with CP approximation for 1:$Y_p=0.5$, 2:$Y_p=0.3$
and 3: $\beta$-equilibrium.}
\label{barras}
\end{figure}

In Fig. \ref{3fig} we display the free energy per baryon for
$Y_p=0.5$ and $Y_p=0.3$. The curves for $\beta$-equilibrium matter
are shown in Fig. \ref{figcomp}.
The three cases show the presence of paste phases, which 
is bigger for larger proton fractions, as already seen in other
works. In Fig. \ref{barras} we can see the pasta structures. For $Y_p=0.5$ 
three different structures are present: droplets (3D), rods (2D) and
slabs (1D), while for $Y_P=0.3$, a small amount of 
tubes (2D) also appear. A similar behavior was obtained in
  \cite{avancini2010} for different models. The reason was pointed out
  to  the non-self consistent treatment of the Coulomb force which
  prevents a redistribution of protons. As a result, the CP method
  predicts smaller extensions of the pasta phases  as a whole and
  for symmetric matter the  larger electron fraction
originates stronger Debye screening effects, and therefore, 
 hinders the appearance of tubes.

The pasta phases shrink with the decrease of
the proton fraction and for $\beta$-equilibrium matter only droplets
are present, as predicted in \cite{oyamatsu2007} for models with a symmetry
  energy slope above 80 MeV at saturation density.
The transition density between the
pasta phases and homogeneous matter shows the same behaviour as in all
models, i.e., it decreases for lower proton fraction and the
lowest value is obtained for matter in $\beta$-equilibrium. The
calculations performed in \cite{Avancini-08} and \cite{Avancini-09} with the CP
method used two different prescriptions for the surface tension
coefficient, based on a fitting of the Thomas-Fermi results to a
Skyrme and to relativistic models respectively. Apart
from these details in the calculations that can modify slightly the quantitative
results, the qualitative conclusions do no differ in general. 

We note that in Fig. \ref{figcomp} (top) the transition core/crust of
a neutron star takes place at $n_B=0.0509$ fm$^{-3}.$ A correlation 
between the transition densities and the slope has been
identified in \cite{Avancini-09,Vidana-09} and in many other works.  
 Since the QMC model has a quite large symmetry energy slope we expect a low crust-core transition density.

\begin{figure}[!]
\begin{tabular}{l}
\includegraphics[width=9.cm,angle=0]{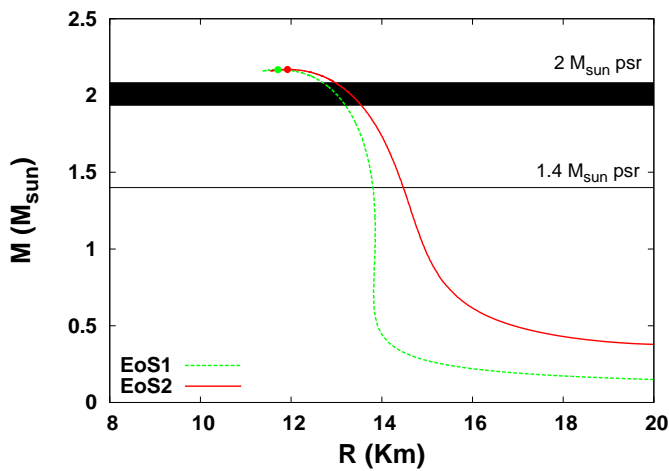}
\end{tabular}
\caption{Mass-radius relation for a family of neutron stars described 
with the QMC model with (EoS1) and without (EoS2) the pasta phases. The thick horizontal 
line represents the $2.01 \pm 0.04$ M$_{sun}$ pulsar PSR J0348+0432 \cite{antoniadis}.}
\label{tov}
\end{figure}

Finally, we analyze the influence of the pasta phases on some neutron
star properties. In Fig. \ref{tov} we show the mass-radius relation. 
The $M(R)$ curves where built with two equations of state: in one of them (EoS1) we consider the occurrence of pasta, whereas in the other (EoS2) no pasta phases are included. We have used the homogeneous QMC EoS for the core, QMC with pasta 
and the Baym-Bethe-Pethick (BBP) \cite{bbp} EoS for the inner crust and the Baym-Pethick-Sutherland (BPS) 
\cite{bps} EoS for the outer crust. In EoS1 the BPS + BBP EoS
goes up to $n_B=1.3 \times 10^{-3}$ fm$^{-3}$, the pasta phases lie in between $n_B=0.15-5.05 \times 10^{-2}$ fm$^{-3}$ when the core EoS takes on. We match the BPS + BBP EoS  
directly to the core EoS for densities
below $8.9\times 10^{-3}$ fm$^{-3}$ for EoS2. Note that the maximum
mass does not change upon the existence of the pasta phases, and both cases reproduce $M_{max}=
2.16$ M$_{sun},$ satisfying the constraints imposed by the recent
measurements  of  the 2$M_{sun}$ pulsars PSR J1614–2230  and PSR J0348+0432 \cite{demorest, antoniadis} 
represented by a horizontal black line in the graphic. 
One of the differences between the results obtained with the two EoS (with and without the pasta phases) appears
when we compare the radius of a typical $1.4$ solar masses 
neutron star. The radius when the pasta phase is included is $13.8
$ km, therefore $600$ m smaller than the $14.4$ km radius obtained
with the BPS+BBP+homogeneous EoS. 
Hence, only the EoS with pasta 
phases is inside the radius range proposed in \cite{HebelerNSR}, where the authors 
constrained the canonical $1.4$ M$_{sun}$ neutron star radii to $R=9.7
- 13.9$ km, or  the  radius range obtained in  \cite{suleimanov16} for X-ray bursting NS.
However, it is outside the range determined in \cite{Ozel:2015fia} from the
analysis of spectroscopic radius measurements during thermonuclear
bursts or in quiescence or in 
\cite{Steiner:2015aea}
from experimental constraints and causality restrictions. In
\cite{helena16a,fortin16} the authors have also shown the sensitivity
of the radius of stars with a mass $\sim 1.4\, M_\odot$ or lower to
the crust EoS and the matching scheme adopted.

\section{Final Remarks}

In the present work we have revisited the calculation of the pasta
phases now using a model with quark degrees of freedom, the QMC model.
The determination of the inhomogeneous phases was possible by
parameterizing the effective nucleon mass as a non-linear function of
the $\sigma$ meson as done before in \cite{guicho96}. 
Part of the results shown in the present work will take part in a  more comprehensive EoS grid that is
being built for star cooling and supernova simulations. 

Our results depend quantitatively on a parameter necessary for the calculation of the surface tensor coefficient. We have
  fitted this parameter to the nuclear surface energy and showed that even changing
  it in a broad interval the pasta extension was only slightly affected.

The general conclusions related to the size of the pasta phases, its
internal structure and the transition density from the pasta to
homogeneous matter go in line with
the ones obtained in previous works \cite{Avancini-08, Avancini-09}.

Calculations that consider $\omega-\rho$ interaction as the ones performed in 
\cite{EPJA_2014,Prafulla_2012,Cavagnoli_2011} are currently under
investigation with the QMC so that its effect on the pasta phase structure
is checked.
We intend to incorporate finite size effects through the implementation of the
CLD prescription \cite{cld_shen, helena15} as well. The CLD presents
amaller discontinuities at very low densities, so it 
can be a useful treatment to obtain all the values that will be needed for a
complete EoS grid.  The inclusion of $\alpha$
particles \cite{avancini10} and other light clusters \cite{avancini12}
can also slightly modify the internal structure of the pasta phases.

\section*{ACKNOWLEDGMENTS}
D.P.M. (grant 300602/2009-0) and G. Grams (doctorate scholarship)
acknowledge support from CNPq and Capes.
C.P. acknowledges partial support from ``NewCompStar'', COST Action MP1304.

\bibliography{biblio.bib}
% \bibliography{alex_database.bib}

\end{document}